\RequirePackage{lineno}
\RequirePackage[2020-02-02]{latexrelease}
\documentclass[prl,aps,amsfonts,amsmath,amssymb,nofootinbib,twocolumn,linenumbers,superscriptaddress]{revtex4}

\usepackage{graphicx}
\usepackage{bm}
\usepackage{hyperref}
\usepackage{natbib}
\usepackage{float}
\usepackage{xcolor}

\restylefloat{table}

\begin{document}

\title{A magnetic continuum observed by terahertz spectroscopy in a quantum spin liquid candidate  BaCo$_2$(AsO$_4$)$_2$}

\author{Xinshu Zhang}

\affiliation{Institute for Quantum Matter, Department of Physics and Astronomy, The Johns Hopkins University, Baltimore, Maryland 21218, USA}

\author{Yuanyuan Xu}
\affiliation{Institute for Quantum Matter, Department of Physics and Astronomy, The Johns Hopkins University, Baltimore, Maryland 21218, USA}

\author{T. Halloran}
\affiliation{Institute for Quantum Matter, Department of Physics and Astronomy, The Johns Hopkins University, Baltimore, Maryland 21218, USA}

\author{Ruidan Zhong}
\affiliation{Department of Chemistry, Princeton University, Princeton, NJ 08544, USA.}

\author{C. Broholm}
\affiliation{Institute for Quantum Matter, Department of Physics and Astronomy, The Johns Hopkins University, Baltimore, Maryland 21218, USA}
\affiliation{NIST Center for Neutron Research, National Institute of Standards and Technology, Gaithersburg, Mary- land 20899, U.S.A}
\affiliation{Department of Materials Science and Engineering, Johns Hopkins University, Baltimore, Maryland 21218, USA}

\author{R. J. Cava}
\affiliation{Department of Chemistry, Princeton University, Princeton, NJ 08544, USA.}

\author{N. Drichko}
\affiliation{Institute for Quantum Matter, Department of Physics and Astronomy, The Johns Hopkins University, Baltimore, Maryland 21218, USA}

\author{N. P. Armitage}
\email{npa@jhu.edu}
\affiliation{Institute for Quantum Matter, Department of Physics and Astronomy, The Johns Hopkins University, Baltimore, Maryland 21218, USA}
 \affiliation{Canadian Institute for Advanced Research, Toronto, Ontario M5G 1Z8, Canada}

\date{\today}

\maketitle

\textbf{Quantum spin liquids (QSLs) are topologically ordered exotic states of matter that host fractionalized excitations.  Kitaev proposed a particular route towards a QSL via strongly bond-dependent interactions on the hexagonal lattice.  A number of candidate Kitaev QSL materials have been pursued, but all have appreciable non-Kitaev interactions, which put these systems far from the QSL regime.  Using time-domain terahertz spectroscopy (TDTS) we observed a broad magnetic continuum over a wide range of temperature and field in the honeycomb cobalt-based magnet,  BaCo$_2$(AsO$_4$)$_2$, which has been proposed to be more ideal versions of a Kitaev QSL.    Applying a small in-plane magnetic field of $\sim$ 0.5 T  suppresses the magnetic order and at at even higher fields gives rise to a spin-polarized state. With 4T magnetic field oriented principally out-of-plane, a broad magnetic continuum was observed that could be consistent with a field induced QSL.  Our results indicate  BaCo$_2$(AsO$_4$)$_2$ to be a promising QSL candidate. }

 \bigskip

 Quantum spin liquids (QSLs) are states of matter where spins are highly correlated and fluctuate quantum mechanically even down to zero temperature~\cite{Balents_Nature, Lee_Science}. In contrast to conventional ordered magnets, which break spin rotational symmetries and can be described by classical order parameters, QSLs are characterized by topological order in the ground state, long range entanglement, and fractionalized excitations~\cite{Broholm_Science}. A prototype QSL was proposed by Anderson in 1973 as the resonating valence bond spin liquid in the triangular lattice~\cite{Anderson1973,Anderson_Science}. The elementary excitations in such QSLs are fractionalized spin 1/2 quasi-particles known as spinons.
 
A milestone in QSLs was the exactly solvable Kitaev model reported by Alexei Kitaev in 2006~\cite{Kitaev_Model}. The Kitaev model consists of S=1/2 spins on a two dimensional honeycomb lattice with bond-dependent Ising-type interactions. The Kitaev Hamiltonian is   $    H_K=\sum_{i,j} K^{\gamma}S_{i}^{\gamma}S_{j}^{\gamma}  $ where $\gamma $=$x,y,z$ denotes the three different bonds and $K$ is the Kitaev interaction constant. By expressing the spin operators $S^{\gamma}$ in terms of Majorana fermions, the ground state is shown to be a QSL with an ensemble of localized and itinerant Majorana fermions~\cite{Kitaev_anyons}.

It has been proposed that transition metal ions in edge-sharing cubic octahedra with strong spin-orbit coupling may be promising candidates for realizing Kitaev QSLs~\cite{Takagi_review}. The main focus for Kitaev QSL candidates has been on $d^5$ pseudospin-1/2 ions with $t^{5}_{2g}$ (S = 1/2, L = 1) electronic configuration, including iridates~\cite{NaIrO3} and the celebrated $\alpha$-RuCl$_3$~\cite{ Banerjee_Nature, Kemp}.  Although these materials order at low temperature,  Kitaev physics may be evinced by the broad continuum at the Brillouin zone center observed in neutron scattering~\cite{RuCl3_Science}.  Magnetic continua as such can be a sign of spin liquid physics as can be indicative of the multi-particle spectra that accompanies fractionalization.  Even more interestingly, applying in-plane magnetic field can suppress the magnetic order and may lead to a field-induced QSL regime~\cite{Zhe_THz, Fieldinduced, continuum_field}. The half quantization of the thermal Hall effect has been observed with a small range of magnetic field, suggesting Majorana edge currents~\cite{ThermalHall,ThermalHall2}.

\begin{figure*}[hbt!]
	\centering. 
	\includegraphics[width=1.9\columnwidth]{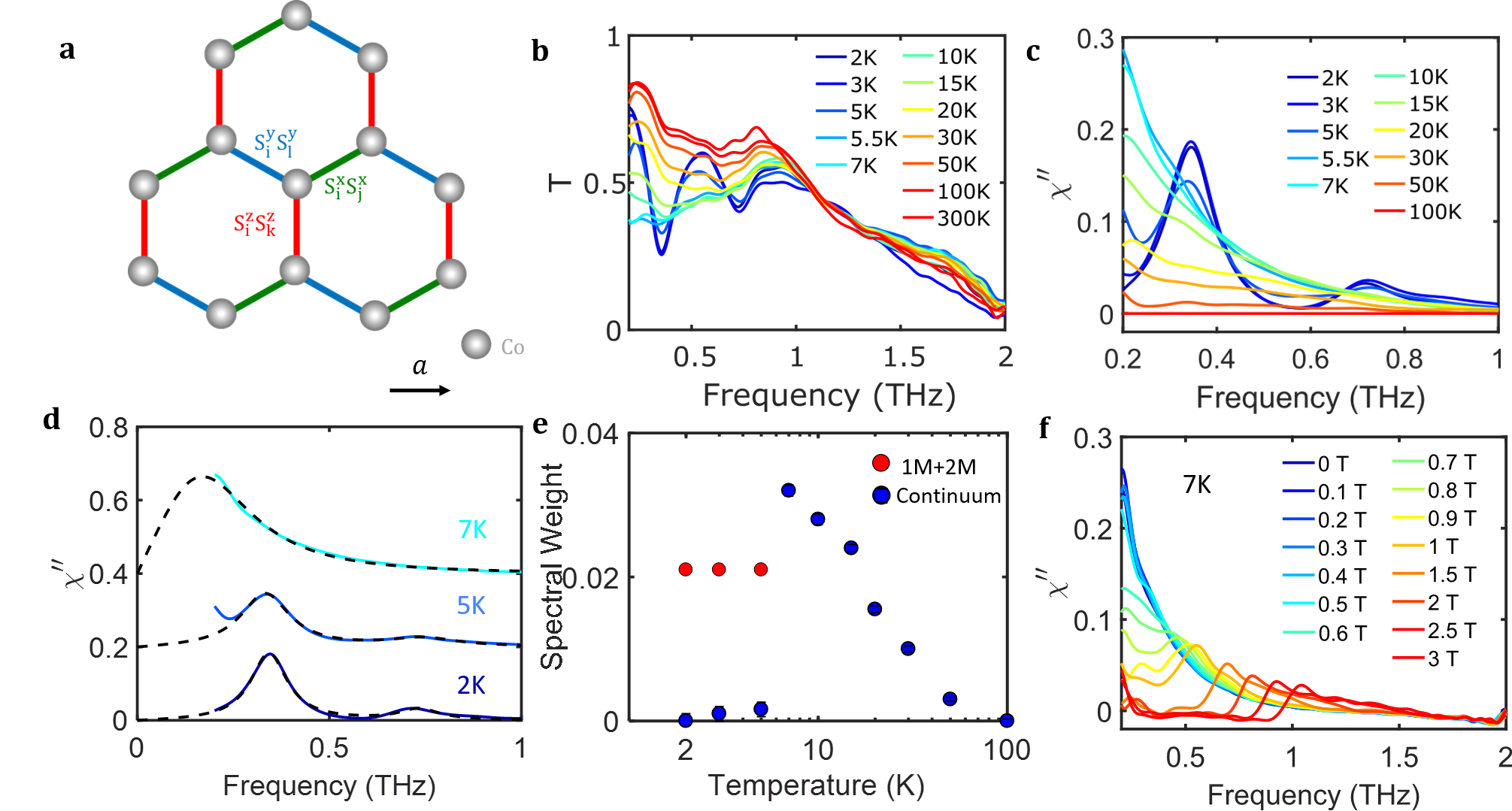}
\caption{   \textbf{(a)} Magnetic Co$^{2+}$ ions form the honeycomb lattice in the ab plane of BaCo$_2$(AsO$_4$)$_2$. Green, blue and red bonds represent the bond-dependent Ising-type Kitaev interactions between two spins on two adjacent sites along $x,y,z$, respectively. \textbf{(b)} Transmission magnitude as a function of frequency at different temperatures from 2 K to 300 K.  Absorptions occur below 1 THz and 100 K, above which the spectra overlap.  \textbf{(c)} The calculated magnetic susceptibility by taking 100K as the reference temperature. \textbf{(d)} Imaginary part of the magnetic susceptibility $\chi''(\nu)$ at 2K, 5K, and 7K.  The Lorentzian fit $\chi''(\nu)=\frac{S\Gamma\nu}{(\nu^{2}-\nu_{0}^{2})^2+\nu^{2}\Gamma^{2}}$ is represented by black dashed lines, where $S$ parametrizes the amplitude, $\Gamma$ is the THz linewidth and $ \nu_{0}$ is the central frequency.  \textbf{(e)}  The temperature dependence of the spectral weight (0.2-1 THz).  Error bars are smaller or comparable to the marker size.  \textbf{(f)}   $\chi''(\nu)$ at 7 K with the in-plane magnetic fields up to 3 T.  }
	\label{fig:1}
\end{figure*}

However, all current Kitaev QSL candidates deviate from the pure Kitaev model due to the presence of significant non-Kitaev terms such as nearest neighbor Heisenberg, off-diagonal exchange, and third neighbor exchanges~\cite{Sears,Sears_prb,Takagi_review} ($J$, $\Gamma$, $\Gamma'$, and $J_3$ terms; see Supplemental Material (SM) for details). The in-plane magnetic field required to destabilize magnetic order is as large as 8 T in $\alpha$-RuCl$_3$ due to these non-idealities~\cite{Zhe_THz, Fieldinduced, continuum_field}. An QSL induced by out-of-plane field has not been observed in $\alpha$-RuCl$_3$, and critical fields as large as 50 T have been proposed~\cite{outplane_theory6}.  Hence, discovering a material that can minimize the non-Kitaev terms is essential. Recent work has predicted that the Kitaev model may be realized in materials based on $d^7$ ions with $t^{5}_{2g}e^{2}_{g}$ (S = 3/2, L = 1) configuration such as Co$^{2+}$, which also have the pseudospin-1/2 magnetic moment~\cite{theory_cobalt, theory_cobalt2, theory_cobalt3, Peter_Kitaev,JeGeunParkReview}. The presence of the extra $e^{2}_{g}$ electrons introduces additional exchange hopping processes and is predicted to suppress the Heisenberg interactions.  Along with other Co-based Kitaev QSL candidate materials~\cite{Vivanco2020,Lin2020a}, BaCo$_2$(AsO$_4$)$_2$ possesses edge-sharing CoO$_6$ octahedra that form a honeycomb lattice~\cite{Ruidan,progressreport,THzCo,theorycolbatates}. A phase diagram consistent with a Kitaev QSL with in-plane field was recently observed in this compound using thermodynamic measurements~\cite{Ruidan}.  In addition to the possibility for smaller non-Kitaev interactions, this material has no stacking faults or twin domains in contrast to $\alpha$-RuCl$_3$~\cite{Ruidan, stackingfaults,Banerjee_Nature} making BaCo$_2$(AsO$_4$)$_2$ an excellent candidate for QSL physics.  However there may be concern that the trigonal distortion present in these and related compounds may quench the orbital moments and suppress the bond-dependence of the exchange couplings.   Moreover, despite the smaller size of the $3d$ orbitals, multiple exchange paths mediated by As may make the third neighbor interactions significant~\cite{theorycolbatates}.  These are not necessarily destabilizing for the QSL, but may change its character and certainly pushing it farther from the exactly solvable Kitaev regime.
\begin{figure*}[t]
	\centering
	\includegraphics[width=1.9\columnwidth]{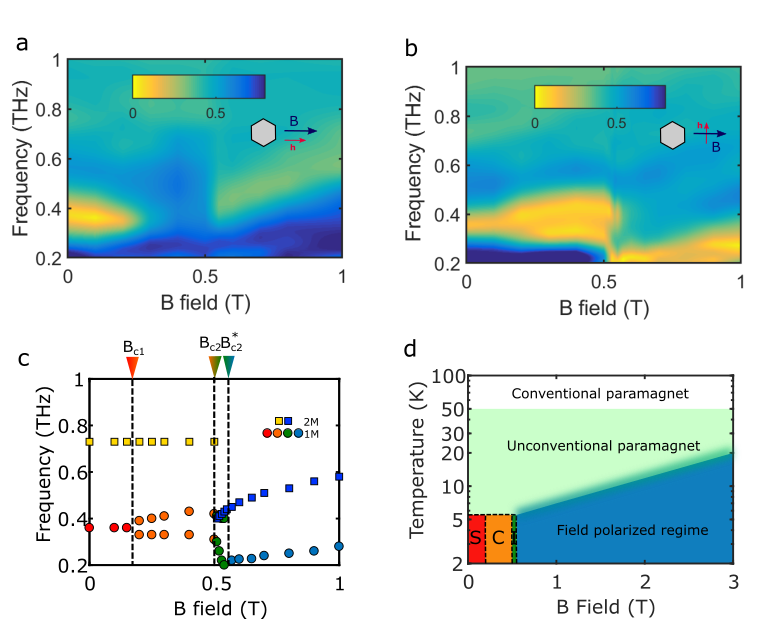}
	\caption{ \textbf{(a)}-\textbf{(b)}  Transmission magnitude as a function of magnetic field and frequency at 2 K in Voigt geometry with $\bf{a}\!\parallel\!\bf{B}\!\parallel\!\bf{h}$ and $\bf{a}\!\parallel\!\bf{B}\!\perp\!\bf{h}$, respectively. The inset indicates the direction of $\bf{B}$ and $\bf{h}$. The data for $\bf{a^{*}}\!\parallel\!\bf{B}\!\parallel\!\bf{h}$ and $\bf{a^{*}}\!\parallel\!\bf{B}\!\perp\!\bf{h}$ is similar and can be found in the SM. \textbf{(c)} Energies of the one and two magnon excitations versus magnetic field along $\bf{a}$ with one and two magnon represented by circles and squares, respectively. Red, orange, green and light blue points represent the one magnon excitations in the spiral ordered state (S), colinear ordered state (C), intermediate state (I) and field polarized regime, respectively. Yellow and dark blue squares represent the two magnon excitations below and above B$_{c2}$, respectively. Three characteristic fields are denoted by inverted triangles with binary colors. The error bars are smaller than marker size. \textbf{(d)} A proposed phase diagram for BaCo$_2$(AsO$_4$)$_2$ with in-plane magnetic field.  The unconventional paramagnet  exhibits a broad magnetic continuum consistent with fractionalized particles.}  
	\label{fig:2}
\end{figure*} 

It has been shown by neutron scattering that Kitaev QSLs exhibit interesting features around the Brillouin zone center~\cite{Banerjee_Nature, RuCl3_Science, Majorana}.  Time-domain terahertz spectroscopy (TDTS) is therefore an important technique as it probes excitations at the Brillouin zone center due to the wavelength of the light much greater than the lattice constants~\cite{Zhe_THz, Liang_THz, Shi_THz, Liang2_THz, Reschke_THz, Sahasrabudhe_THz}. The growth of the crystals used in these measurements is described in Ref.~\cite{Ruidan}. The absence of symmetry forbidden Raman phonons and the narrow linewidths of the symmetry allowed ones are testament to the high crystal quality (see SM).  Fig.~~\ref{fig:1}(b) shows the THz transmission magnitude at zero magnetic field from 2 K to 300 K.  Dissipative features occur below 1 THz and below 100 K. We can obtain the magnetic susceptibility (Fig.~~\ref{fig:1}(c)) from the transmission at each temperature by referencing to the high temperature transmission~\cite{Xinshu} (Methods).  It is clear that a broad continuum gradually increases upon cooling until the Néel temperature $T_{N}$ $\sim$ 5.4 K, below which the continuum is replaced by a spectra with sharp peaks. The observation of a continuum at the $\Gamma$ point at finite T is consistent with a so-called Kitaev paramagnet and the existence of fractionalized spin excitations e.g. Majorana excitations~\cite{Banerjee_Nature, RuCl3_Science, Majorana, continuum_theory, continuum_theory2,Raman_RuCl3}.

Below $T_{N}$, excitations develop that we assign to one magnon (1M) and two magnon (2M) character (centered at 0.35 and 0.75 THz).  The 1M feature is one of the two expected for ordered spins on the hexagonal lattice.  1M and 2M excitations were assigned based on the selection rules discussed below.  The one and two magnon excitations can be fit by Lorentzian functions to determine their spectral weight (See caption Fig.~\ref{fig:1} and SM). It is interesting to note that, despite the $T_{N}$ $\sim$ 5.4 K, at 5 K the continuum is still visible, but that it disappears by 2 K.  A sum rule analysis based on Kramers-Kronig
relations relates the  $\chi''(\nu)$ to the dc susceptibility $\chi'(0)$ 

\begin{equation}
 \chi'(0)= \frac{2}{\pi}\int_{0}^{\infty} \frac{\chi'' (\nu)}{\nu} d\nu. 
\label{eq:6.7}
\end{equation}  
At 2 K, the obtained dc contribution from the fits to the one and two magnon peaks
is $\sim$ 0.07(1), which is much smaller than the value $\sim$ 0.2 determined from dc susceptibility measurement. Hence, there must be some remaining continuum below our measurement range even down to 2 K. The coexistence of the continuum and magnons over even a small temperature range indicates an unconventional ordered state as all spectral weight would usually be in the magnon excitations in a conventional magnet.  Similar features were also observed in $\alpha$-RuCl$_3$~\cite{Banerjee_Nature, RuCl3_Science,Liang_THz}. The continuous suppression of the continuum below $T_{N}$ is consistent with fractionalized particles in a spin liquid-like model being replaced by magnons in the long-rang ordered state.

Above $T_{N}$, the continuum is stable with increasing temperature and persists almost unchanged up to 50 K e.g. $\sim$ 10 $T_{N}$. This temperature range may be the so-called Kitaev paramagnet regime, where the physics is not particularly sensitive to thermal fluctuations~\cite{Majorana, continuum_theory2}. By fitting the real and imaginary magnetic susceptibility to a Lorentzian form simultaneously along with the constraint of dc susceptibility, we are able to capture the broad continuum down to zero frequency as shown in Fig.~~\ref{fig:1}(d) (SM). The broad continuum is unlikely to be due to paramagnons given the center of the continuum at finite frequency as shown in Fig.~~\ref{fig:1}(d).  Moreover,  generally TDTS (a \textbf{q}=0 probe) can not generally resolve paramagnons as one must typically probe them at a wavevector much longer than the inverse correlation length of the magnetic order.   This is essentially impossible with a \textbf{q}=0 probe.   Moreover, paramagnons usually exhibit strongly temperature (and/or field) dependent spectra, which is inconsistent with the weak temperature and field dependence of the continuum.  It is noteworthy that the shape of the present continuum is distinct from the continuum in $\alpha$-RuCl$_3$, which has a long tail at higher frequency and was interpreted as magnon decays ~\cite{Banerjee_Nature, RuCl3_Science,magnondecay,magnondecay2}. However, we note that the shape of the continuum in BaCo$_2$(AsO$_4$)$_2$ is more consistent with the shape of predicted multi-particle continuum with dominant Kitaev terms ~\cite{continuum_theory, continuum_theory2,magnondecay}.  In theory~\cite{continuum_theory}, the continuum signal is prominent only up to about 25$\%$ of the bandwidth (giving the total magnetic bandwidth $W \approx 16$ meV).  At even higher temperature $\sim$100 K, this anomalous physics fades away and the material enters a conventional paramagnetic regime.

The continuum can also be affected by applying magnetic field. Applying in-plane magnetic field at 7 K converts the continuum to conventional magnons similar to entering the ordered state.   This is consistent with the spin-liquid like paramagnetic state being connected to the field polarized regime smoothly as shown in Fig.~~\ref{fig:1}(f). The continuum starts decreasing its intensity at 0.6 T and the two magnon excitation develops and shifts to higher frequency.  

\begin{figure*}[hbt!]
	\includegraphics[width=1.9\columnwidth]{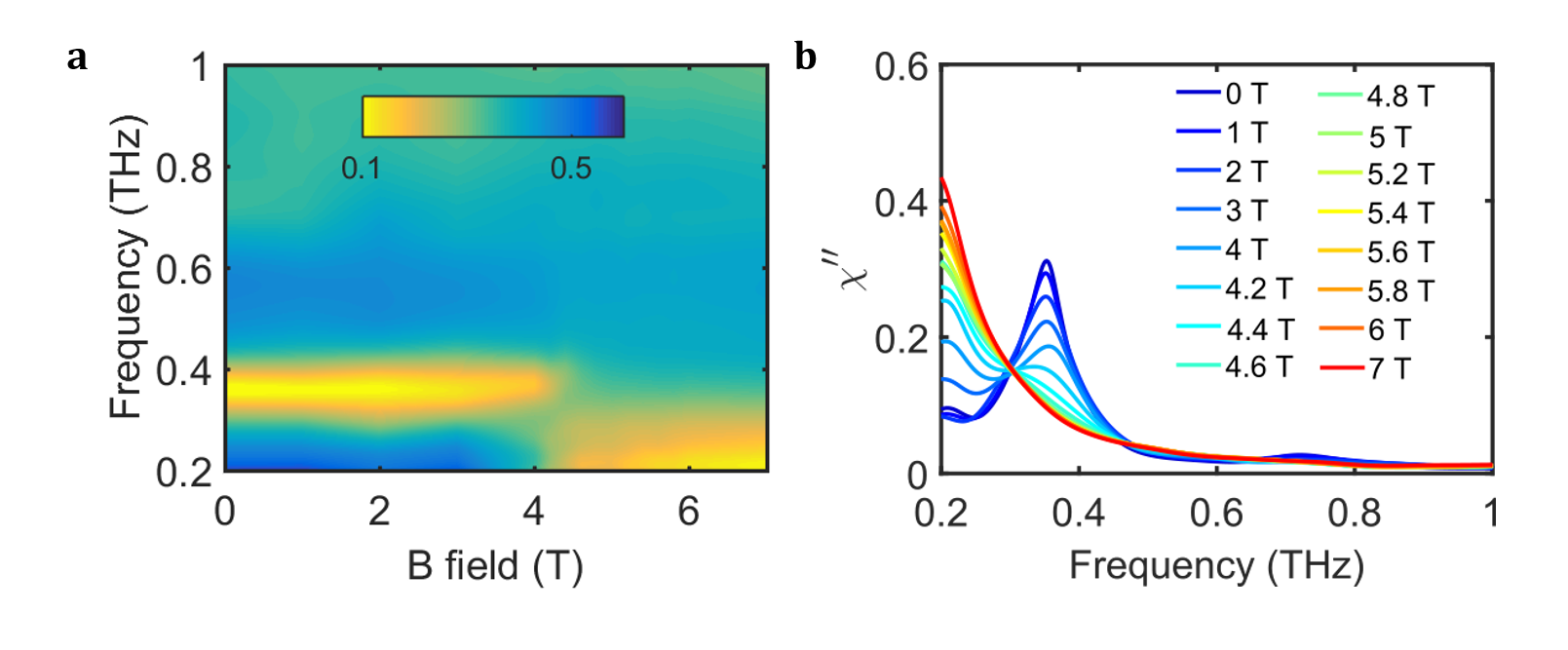}
	\caption{\textbf{(a)} Color plot of the transmission magnitude as a function of magnetic field and frequency at 5 K in Faraday geometry with out-of-plane magnetic field. \textbf{(b)} Imaginary magnetic susceptibility $\chi''(\nu)$ at 5K  from 0 T to 7 T. }
	\label{fig:3}
\end{figure*} 

Next, we explore the effects of applying an in-plane magnetic field on the low temperature state (Voigt geometry ($\bf{k}\!\perp\!\bf{B}$), where $\bf{k}$ is the THz wavevector) at low temperatures.  Fig.~~\ref{fig:2}(a) and (b) show the color plots of the transmission magnitude as a function of magnetic field and frequency at 2 K with $\bf{B}\!\parallel\!\bf{h}$ and $\bf{B}\!\perp\!\bf{h}$, respectively.  $\bf{B}$ is the static magnetic field and $\bf{h}$ is the THz ac magnetic field. $\bf{B}\!\parallel\!\bf{h}$ and $\bf{B}\!\perp\!\bf{h}$ have distinct optical selection rules for one and two magnon excitations. One and two magnon excitations were distinguished by their energy and polarization dependence.  $\bf{B}\!\parallel\!\bf{h}$ is the so-called ``longitudinal” configuration,  where THz mainly excites two magnon excitations unless the spins are not perfectly aligned with external magnetic field B (i.e. below 0.2 T). In the ``transverse" configuration $\bf{B}\!\perp\!\bf{h}$, THz mainly excites one magnon excitations and is less sensitive to the two magnon feature. 

By tracking changes to the excitation spectrum, we can reconstruct the phase diagram for in-plane field.  By comparing Fig.~~\ref{fig:2}(a) and (b), one can identify  three characteristic fields B$_{c1}\sim$0.2 T, B$_{c2}\sim$0.5 T and B$_{c2}^{*}\sim$0.55 T. The one magnon and two magnon energy versus magnetic field obtained from both configurations are summarized in Fig.~~\ref{fig:2}(c). Below B$_{c1}$, BaCo$_2$(AsO$_4$)$_2$ has been reported to be in a spiral long range ordered state~\cite{earlyneutron, earlyneutron2, Ruidan}. The spins in the spiral have components transverse to THz magnetic field $\bf{h}$, therefore one magnon excitations can be observed in both $\bf{B}\!\parallel\!\bf{h}$ and $\bf{B}\!\perp\!\bf{h}$ configuration. As B$_{c1}<$ B $<$ B$_{c2}$, it goes into a colinear ordered state with 1/3 plateau observed in magnetization measurements. The spins are orientated along the magnetic field, hence no magnon peaks are observed in $\bf{B}\!\parallel\!\bf{h}$. In the colinear phase, the two spins in the unit cell point in opposite directions, giving rise to a splitting of the high energy spin wave mode as observed in $\bf{B}\!\perp\!\bf{h}$ and represented by orange points.

It is unclear at this time if region from B$_{c2}<$ B $<$B$_{c2}^{*}$, constitutes a distinct phase.  The intensity of the upper one magnon branch is reduced abruptly with almost unchanged energy, while the energy of the lower one magnon branch drops quickly but smoothly below our range as shown in  Fig.~~\ref{fig:2}(b).  These critical fields are consistent with previous magnetization and ac susceptibility measurements~\cite{Ruidan}.  The abrupt suppression of the magnon excitations may imply a field-induced intermediate state in the field range of B$_{c2}<$ B $<$B$_{c2}^{*}$. Although magnetization data only shows a single feature in this range, we note that similar features have been observed and interpreted as a signature of field-induced spin liquid in $\alpha$-RuCl$_3$~\cite{Zhe_THz} for the field ranges that show the quantized thermal conductance~\cite{ThermalHall,ThermalHall2}. Nevertheless from our data, we cannot not exclude the possibility of this region being not a phase, but a transition region; further experiments such as thermal conductance measurements~\cite{ThermalHall,ThermalHall2,oscillations} will be helpful to provide insights.  Above B$_{c2}^{*}$,  the system enters into a field polarized regime.   Again two features appear.  A two magnon excitation is apparent at 0.4 THz and moves to higher frequency with increasing field, while one magnon excitations have half the energy.

A phase diagram of BaCo$_2$(AsO$_4$)$_2$ with in-plane magnetic field can be obtained as shown in Fig.~~\ref{fig:2}(d). The full spectra up to 4 T are shown in the SM.  Below $T_{N} \sim$5.4 K, upon applying field, BaCo$_2$(AsO$_4$)$_2$ experiences a phase transition from the spiral ordered state (S) to colinear ordered state (C) at B$_{c1}\sim$0.2 T, followed by a {\it transition region} for  B$_{c2}<$ B $<$B$_{c2}^{*}\sim$0.55 T and ultimately into the field polarized phase. In the intermediate temperature regime, BaCo$_2$(AsO$_4$)$_2$ is an unconventional paramagnet characterized by a broad continuum reminiscent of a spin liquid. It is connected continuously to the field polarized regime and conventional paramagnetic state with applying field and increasing temperature, respectively.

In a Kitaev magnet the critical fields required to suppress the magnetic ordered state and induce a QSL regime depends on the strength of the non-Kitaev interactions~\cite{RuCl3_exchange2,outplane_theory1,rethinking_RuCl3}. The critical fields B$_{c2}$, B$_{c2}^{*}$ in BaCo$_2$(AsO$_4$)$_2$ are 15 times smaller than similar ones in $\alpha$-RuCl$_3$, indicating that if a Kitaev model applies it has much smaller non-Kitaev interactions in BaCo$_2$(AsO$_4$)$_2$.   Moreover, as discussed in Ref.~\cite{rethinking_RuCl3} and the SM one can estimate from the field dependence of the lowest energy magnon mode in the paramagnetic regime the combination of the symmetric off-diagonal exchanges $\Gamma_{total} = \Gamma + 2\Gamma'$.   From such an analysis we can put a bound on $\Gamma_{total} \approx $ 3 meV) (or alternatively within the XXZ-$J_{\pm\pm}$-$J_{z\pm}$ model this is a bound on the quantity $\Delta J_1 - J_1 $; see SM).  This is about a factor of 3 smaller than estimates for this parameter for $\alpha$-RuCl$_3$~\cite{rethinking_RuCl3} and is in good agreement with theory~\cite{theory_cobalt, theory_cobalt2}. It is also a smaller fraction of the total magnetic bandwidth ($W = 10-16$ meV as estimated from the  continuum discussed before or from the neutron scattering dispersion in field on a related material~\cite{Nair}) than it is in $\alpha-$RuCl$_3$~\cite{Banerjee_Nature, RuCl3_exchange, RuCl3_exchange2}.   Note that while we find  $\Gamma_{total}$ relatively small, it is not insignificant as otherwise the relative anisotropy for in- and out-of-plane field would not be as large as observed.

Finally (and perhaps most importantly) we give evidence that an out-of-plane field can realize a more stable QSL state as compared to in-plane field as proposed theoretically~\cite{outplane_theory1,outplane_theory2, outplane_theory3, outplane_theory4,outplane_theory5, scaleinvariant}.  Fig.~~\ref{fig:3}(a) shows the transmission magnitude at 5 K as a function of out-of-plane magnetic field and frequency in the Faraday geometry ($\bf{k}\!\parallel\!\bf{B}$). The sharp magnon excitations are suppressed completely at $\sim$ 4.5 T. Above 4.5 T,  a continuum appears and persists up to the highest field we can apply (7 T). This is consistent with the formation of a multi-particle continuum and a field-induced QSL regime being stabilized up to at least 7 T before entering a field polarized regime. The imaginary magnetic susceptibility  $\chi''(\nu)$ is shown in Fig.~~\ref{fig:3}(b). It is clear the continuum is distinct from the conventional spin-wave excitations that would exist in a field polarized paramagnet.  Furthermore, we observe the continuum in both left and right circular basis (SM). This gives further evidence against the possibility of the field polarized regime as spin waves are typically observed only in one circular basis and not the other due to optical selection rules~\cite{Xinshu}. The feature's relative stability with field and the \textbf{q}=0 nature of our probe again means it is unlikely that it arises via sources like paramagnons or short range correlation for the same reasons discussed above for the finite temperature case.   Magnon decay is also unlikely as magnon damping would be expected to decrease at higher fields where the zone-center magnons and multiple magnon continuum will overlap less~\cite{magnondecay,magnondecay2}.

Note that the appearance of a continuum with out-of-plane field has not been observed in $\alpha$-RuCl$_3$ or other Kitaev QSL candidates presumably due to the very large magnetic fields ($>$ 50 T) that are needed to suppress their ordered state~\cite{outplane_theory6} (due to their larger non-Kitaev exchanges).   Also note, that it is likely that there is a small in-plane field that exists for the data in Fig. 3, as ac susceptibility and magnetization measurements (SM) show that the transition for out-of-plane field may actually be as high as 11 T.  It is likely that the critical field for primarily out-of-plane field is exceedingly sensitive to a small in-plane field as the in-plane critical field is very small and the spins are easy-plane.  Future susceptibility experiments with a rotation stage are required to determine the transition field more accurately.  Nevertheless, note that despite the fact that a very small in-plane field may exist for the data in Fig. 3, the observed field-induced continuum is not simply related to an in-plane field as no continuum was been observed for $\bf{k}\!\perp\!\bf{B}$. Also note that the continuum with primary out-of-plane field was observed at 4K around 5.5T but not observed at 2K up to 7 T, which presumably requires larger field (SM). Although the continuum was observed at temperatures close to $T_{N}$, as discussed it cannot be characterized as a conventional paramagnet close to an ordered state.    

We have reported a time-domain terahertz spectroscopy study of the cobalt based Kitaev QSL candidate BaCo$_2$(AsO$_4$)$_2$. At zero field, we find a broad continuum consistent with the existence of fractionalized particles. The response of the continuum to temperature and magnetic field are also consistent with the predictions of the Kitaev model.  Applying a very small in-plane magnetic field of $\sim$0.5 T  suppresses the magnetic order, suggesting relatively small non-Kitaev terms. Importantly,  a broad continuum was observed with a primary out-of-plane magnetic field that could be the evidence of  the field-induced QSL state.   Despite this broad agreement, we note that we have not explicitly observed evidence for the bond-dependent Kitaev exchanges.  The role of trigonal distortion should be investigated, as large enough distortions will quench the orbital degrees of freedom necessary for such exchanges.   Moreover, we cannot exclude the important role of third-nearest neighbor interactions~\cite{theorycolbatates}.  Irrespective of these open questions, we hope our work may stimulate further study of Co$^{+2}$ Kitaev QSL candidates  as well as bring to bear other powerful probes like neutron scattering.



\subsection{Time-domain terahertz spectroscopy}

Time-domain THz spectroscopy was performed in a home built system at Johns Hopkins University. THz pulses with a bandwidth between 0.2 to 2 THz were
generated by a photoconductive antenna Auston switch (emitter) upon illumination by an infrared laser and then detected by another Auston switch (receiver). We measured a single crystal with $2 \times 2\times 0.5$ mm and an aperture with the same size as an reference.  The electric field profiles of the THz pulses transmitted through the sample and an identical bare aperture were recorded as a function of time by moving a delay stage and then converted to the frequency domain by Fast Fourier Transforms (FFTs). By dividing the FFTs of the sample and aperture scans, we obtain the complex transmission of the sample.

\subsection{THz magnetic susceptibility}

The complex transmission  is given by the relation $\tilde{T}(\nu)=[4\tilde{n}/(\tilde{n}+1)^{2}]\mathrm{exp}[i2\pi\nu d(\tilde{n}-1)/c]$, where $\tilde{n}=\sqrt{\tilde{\epsilon}\tilde{\mu}}$ is the complex index of refraction, $\tilde{\epsilon}$ is the dielectric constant, $\tilde{\mu}=1+\tilde{\chi}$, $c$ is the speed of light and $d$ is the sample thickness. We determine $\tilde{n}$ at each temperature using the Newton-Raphson method and then isolate $\tilde{\chi}$ by measuring the sample at a reference temperature above which the spectrum does not change any more. At this reference temperature 100 K in this case,  $\tilde{\chi}_{ref}$ can be taken to be zero  so $\tilde{n}_{ref}=\sqrt{\tilde{\epsilon}}$. For the low temperature, $\tilde{n}_{low}=\sqrt{\tilde{\epsilon}(1+\tilde{\chi}_{low})}$. Thus, the low temperature magnetic susceptibility is given by $\tilde{\chi}_{low}=(\tilde{n}_{low}/ \tilde{n}_{ref})^{2}-1$ .

 \section{Acknowledgements:  }
 
This was supported as part of the Institute for Quantum Matter, an EFRC funded by the DOE BES under DE-SC0019331.  NPA had additional support from the Quantum Materials program at the Canadian Institute for Advanced Research.  We would like to thank P. Chauhan and A. Legros for critical comments on this manuscript and H.-Y. Kee, G. Khaliullin, and H. Liu for helpful conversations.
 
 \section{Author contributions:  }
 
XZ performed THz experiments and analyzed the data. RZ and RC grew the single crystals. YX and ND performed Raman spectroscopy. TH and CB performed magnetization experiments.  XZ and NPA prepared the first draft, and all authors contributed to writing the manuscript.

 \section{Additional information:}

\textbf{Competing financial interests:} The authors declare no competing financial interests.
 
 \section{Data availability}
 The data that support the findings of this study are available from the corresponding authors upon reasonable request.

\normalsize

\end{document}